\documentclass[12pt]{article}

\newcommand{\beq}{\begin{equation}}
\newcommand{\eeq}{\end{equation}}
\newcommand{\bea}{\begin{eqnarray}}
\newcommand{\eea}{\end{eqnarray}}

\newcommand{\ket}{\rangle}

\begin{document}

\begin{center}
{\Large Implications for cognitive quantum computation and decoherence
limits in the presence of large extra dimensions}
\vskip 0.5cm
J. R. Mureika\\
Department of Physics, Loyola Marymount University\\
1 LMU Drive, Los Angeles, CA~~90045-8227
\vskip .2cm
email: jmureika@lmu.edu
\end{center}

\vskip 1cm

\noindent{\bf Abstract}\\
An interdisciplinary physical theory of emergent consciousness has previously 
been proposed, stemming from quantum computation-like behavior between 
$10^9$ or more entangled molecular qubit states (microtubulin).  This 
model relies on the Penrose-Di\'osi gravity-driven wavefunction collapse 
framework, and thus is subject to any secondary classical and quantum 
gravity effects.
Specifically, if large extra spatial dimensions exist in the Universe,
then the resulting corrections to Newtonian gravity cause this model 
to suffer serious difficulties.  It is shown that if the extra dimensions are
larger than 100~fm in size, then this model of consciousness is unphysical.
If the dimensions are on the order of 10~fm in size, then a significantly 
smaller number of microtubulin than originally predicted are required to 
satisfy experimental constraints.  Some speculation on evolution of 
consciousness is also offered, based on the possibility that the size of these
extra dimensions may have been changing over the history of the Universe.\\

\pagebreak

\section{Introduction}
Consciousness as an emergent phenomenon of physical and/or biological systems
is a growing field of interdisciplinary interest.  Traditionally, biophysical
and neuroscientific descriptions of brain processes have been restricted to
classical neural network designs 
(see {\it e.g.} \cite{arbib} and references therein).  However, there is 
increasing evidence to suggest that many biological processes rely on
quantum mechanical behavior, including protein folding \cite{protein},
single-photon activation of the rhodopsin {\it cis-trans} isomers in
visual pre-perception \cite{rhodop}, and neuron impulse transmission across the 
synaptic gap \cite{neuron}.

The quantum computing revolution \cite{qcomp} has spawned a new 
frame of reference from which to view the problem of consciousness.  
To reconcile the aforementioned ``bio-quantum mechanical'' phenomena
with the common interpretation of the brain as a computational device, recent
proposals have sought to describe higher cognitive functions as
quantum processes.  Comprehensive summaries of this 
burgeoning field can be found in Reference~\cite{tusz}, but specific
important contributions to the field include \cite{lockwood,stapp,page,
eccles,hodgson,orchor}.  From a slightly different perspective,
Davies \cite{davies} has proposed a link between biological complexity and 
holographic theories of cosmological entropy bounds.

Classical computation is based on a binary system of bits which 
must either be in one state (1) or the other (0).  Memory storage and
computation proceed by assigning values to a string of bits and performing
logical operations in series.  Such a ``linear'' processing system 
ultimately limits the speed with which calculations can be performed.
Quantum computers export the superposition principle of states, replacing
the classical bit with a two-state quantum bit (qubit) $|\psi\ket$, which
can assume both values (0 and 1) simultaneously.  Quantum logic operations
are performed on the combination of qubits in the product state
$|\psi_1\ket \otimes |\psi_2\ket \otimes |\psi_3\ket \otimes \cdots$,
which force evolution to an entangled superposition
of states.  Since this enables the quantum computer to evaluate multiple
and simultaneous ``solutions'' to the logical operations,
massively-parallel computations can be effected.

Among the many immediate benefits of quantum computing is the 
decreased computation time required for classically-difficult problems. 
This includes prime number factorization algorithms that execute in 
polynomial time as opposed to exponential time \cite{shor} (as a function
of the size of the number), as well as  
fast database search algorithms \cite{grover}.  
Both of these are particularly attractive 
features to cognitive scientists, since the  classical computational time
required to perform equivalent tasks is astronomical.

The primary obstacle to a realization of quantum computation is preventing
the qubit entanglement from decohering via interactions with the surrounding 
environment.  This is generally done by isolating the system from the
environment, and reducing the temperature to exceedingly low values, 
making the notion of room-temperature (and desktop) 
quantum computers a distant reality.

Such rapid decoherence is arguably the death-knell for most
quantum theories of cognition.  The academic community is split on whether
or not this is an obstacle that Nature has managed to overcome in designing
a biological quantum computer (see the discussion in \cite{tegmark}).
The most developed theory of cognitive quantum computation -- dubbed the
``Orchestrated Objective Reduction'' (Orch-OR) mechanism -- has been 
proposed by Penrose and Hameroff \cite{orchor}, who argue that
environmentally-isolated quantum entanglement states are kept coherent 
long enough to perform conscious processes.  

However, the Orch-OR mechanism relies on 
classical Newtonian gravity to address a largely quantum mechanical issue.  
If a theory wishes to combine gravitation with quantum effects,
then {\it all} aspects of quantum gravity should be addressed.  In particular,
the accuracy of Newtonian gravity at small (sub-micron) scales has recently 
come under scrutiny, thanks to string-inspired theories which propose the 
existence of large extra spatial dimensions \cite{add,rs}.  

This paper will thus re-examine the feasibility of the Orch-OR in light of
the possible existence of large extra spatial dimensions (in the Arkani-Hamed,
Dvali, Dimopoulos model).  First, a more detailed synopsis of the
debate surrounding the Orch-OR model are discussed in Section~\ref{debate}.
The physical basis for the gravity-driven collapse mechanism is 
reviewed in Section~\ref{ormechanism}, and a brief introduction to large
extra dimensions follows in Section~\ref{ledsection}.  Associated modifications
to the gravity-collapse model are discussed in Section~\ref{ledcollapse},
including the potential impact on ``traditional'' quantum mechanical 
phenomena such as nucleon superposition.
Finally, the impact on Orch-OR is addressed in Section~\ref{orchorled}.
It is shown that the decoherence times calculated in \cite{orchor}
are greatly affected by such a supposition.  Should extra dimensions
exist and be of sufficiently large scale, the Orch-OR scenario could 
suffer serious setbacks.

Before proceeding, it should be emphasized from the outset that
the analysis herein is based on a combination of both hypothetical
and arguably untraditional models (Orch-OR), as well as accepted but untested 
theory (large extra dimensions).  
For the sake of this argument, it is assumed that the Orch-OR model is 
correct.  The cautious reader should thus approach the paper as an exercise 
in academic discourse and open-minded speculation.

\section{A summary of the debate: Orch-OR or not Orch-OR?}
\label{debate}
The ``Orchestrated Objective Reduction'' (Orch-OR) mechanism 
posits that conscious 
``computation'' does not take place in the classical neural circuitry of 
the brain, but rather in the constituent molecules (microtubules) of the 
cellular cytoskeleton.  Microtubules are hollow, cylindrical structures whose
walls consist of 13 chains of the protein tubulin. These proteins can 
assume two 
distinct physical conformations resulting from different electric dipole
moments along their physical axis.  Consequently, this ``two-state'' behavior 
suggests that 
the tubulin protein is a prime candidate for a qubit.  Penrose \cite{orchor} 
has proposed that each tubulin qubit can become entangled with other local 
tubulin to form a superposition state.
These superpositions are unstable and subject to collapse, and the mechanism
which drives this collapse has been proposed to be that of Penrose
\cite{penrose1,penrose2,penrose3}, coined ``objective reduction''.

The primary criticism of this model is that the brain is not an isolated 
low temperature environment, and thus the decoherence of any macroscopic 
entanglements would be effectively instantaneous ($\sim 10^{-20}~$sec) 
due to other local quantum processes such as ion interactions \cite{tegmark}.  
This criticism is countered by arguing 
that the microtubules are sufficient shielded by molecular and electrostatic 
considerations,
thus the entangled states can survive for macroscopic time intervals 
\cite{qcmicro}.

Furthermore, the authors of \cite{orchor} suggest that well-known 
electrophysiological ``brain-wave'' frequencies are signatures of the extended 
superposition collapse.  For example, they demonstrate that the Orch-OR 
mechanism allows tubulin qubit superpositions to be maintained for as long as 
25~microseconds, corresponding to the 40~Hz thalamo-cortical coherent 
oscillation frequency.  This association (among others) has been offered as 
evidence in support of the quantum computing model for consciousness.  

In the Orch-OR model, the culprit for the wavefunction collapse is gravity.
Penrose \cite{penrose1,penrose2} has postulated that gravitational interactions
between the different {\it physical} configuration of each qubit eigenstate
introduces a type of time-energy uncertainty relationship.  This ultimately
limits the duration that a state can remain in superposition, and is
discussed in detail in the next section.

\section{The physical foundations of Objective Reduction}
\label{ormechanism}
The Objective Reduction (OR) mechanism is a potential solution to the
measurement problem between the two (or more) possible eigenstates 
of an evolving wavefunction, $|\Psi\rangle = 
\alpha |\psi_1\rangle + \beta |\psi_2\rangle$.  
A complete derivation of the OR mechanism will not be reproduced here, but
can be found in References~\cite{penrose1,penrose2,penrose3}.
Similar collapse schemes have been proposed by other authors
\cite{diosi,ghirardi,pearle,tamas}, but will not be discussed in the
present context\footnote{It should be pointed out that the mechanism proposed in \cite{diosi}
predates and ultimately produces similar results to Penrose's formalism.}.
A review of four distinct interpretations of such collapse mechanisms
may be found in \cite{diosinew}.

The theory proposes that each eigenstate in the superposition possesses
a conformationally-distinct physical orientation.  The overlap of both
conformations will have a small but relevant impact on the local curvature
of spacetime.  The net result is that there will be two distinct spacetime
curvatures in superposition with one another.  Each curvature can be represented
by a quantum state $|{\cal G}_i\rangle$, correlated with the eigenstate
$|\psi_i\rangle$, and thus the actual particle 
wavefunction will be represented by the entanglement
$|\psi_1\rangle |{\cal G}_1\rangle + \beta |\psi_2\rangle |{\cal G}_2\rangle$.
The geometric superposition creates an ill-definition of the time-like 
Killing vector, $\partial/\partial t$, which will ultimately lead to the
collapse.  Simply put, each eigenstate will follow its own unique free-fall
vector, simultaneously violating the weak equivalence principle.
                                                                                
It can be shown that this ``instability'' in the superposition is limited
by an upper bound on the gravitational interaction energy between the
two eigenstate conformation states.  Assuming the separation of each state 
($\Delta r$) exceeds their own physical extent, this is simply the Newtonian 
energy
\beq
E_{\Delta} \sim \frac{Gm^2}{\Delta r}~,
\label{lifetime}
\eeq
The states can just as easily overlap physically, although the calculation
becomes more complicated.
The collapse time of the spacetime superposition is determined by
the uncertainty relation
\beq
T_c \sim \frac{\hbar}{E_\Delta}~~,
\label{tcpen}
\eeq
and thus is inversely proportional to the gravitation energy of the system.
This result is a consequence of the framework in \cite{diosi},
as well as the alternate derivation in \cite{penrose1,penrose2,penrose3},
and the interested reader is directed to these References for a complete
review of the foundations.

Hence, in this scheme
a nucleon of mass $10^{-27}~$kg whose superpositions are separated by the
strong interaction scale of $10^{-15}$~m will remain superposed for
$T_c \sim
10^{15}~$seconds (or about $10^7$~years), whereas a possible macroscopic
superposition having larger $E_{\Delta}$ will decay relatively quickly.
For example, if a speck of dust ($m \sim 10^{-6}~kg$) evolves into 
superposed states that are separated by 0.01~mm, the wavefunction would 
collapse in well under $10^{-16}$~seconds.

%
%

\section{Orchestrated Objective Reduction}
\label{penham}
Orchestrated Objective Reduction (Orch-OR) is the application of the
OR formalism to superpositions of entangled tubulin ``qubits'' in the
cellular cytoskeleton.  
Using this rationale, the authors of \cite{orchor} have calculated a 
variety of constraints
on the nature of tubulin structures in which conscious quantum computations
could be realized.  Various electrophysiological frequencies are 
known to exist in the brain -- most notably the 40~Hz thalamo-cortical coherent 
oscillations ($\Delta t = 25$~ms), 10~Hz alpha rhythm EEG ($\Delta t =100~$ms), 
and Libet's pre-conscious 2~Hz sensory threshold ($\Delta t = 500~$ms) 
\cite{orchor}.  If this period $\Delta t$
corresponds to the collapse time for tubulin qubit entanglements, then
it can be reasoned that
$\Delta E \sim \frac{\hbar}{\Delta t} \approx 10^{-15}~{\rm eV}$
is the required gravitational self-energy which must be attained by 
the system for the longest period of 500~ms.
This implies that $N$ qubits possessing an OR energy $E_0$ must be entangled
together, where $\Delta E = NE_0$.  

In order to estimate the self-energy $E_0$ of one tubulin superposition, 
it has been suggested \cite{orchor} that the most appropriate conformation
states are those in which a carbon-12 nucleus is superposed within its own
atomic radius ($a \sim 2.5 \times 10^{-15}~$meters).  In this case, the 
self-energy term becomes
$E_0 = \frac{Gm_C^2}{a} \sim 10^{-28}~{\rm eV}$,
and thus $N = \Delta E/E_0 \sim 10^{13}$~carbon atoms, or 
$10^{14}$ nucleons.  Since each microtubule is composed of 
$\mu \sim 10^5$ nucleons, 
this implies that $N_T \sim 10^9$ tubulin are required for a conscious 
instance.  This corresponds to roughly 100 neurons, each containing $10^7$
tubulin \cite{yubaas}.

Combining these into one single expression, it can be concluded that the
number of microtubulin $N_T$ required to form a $\Delta t$~second 
pre-conscious instance is
\beq
N_T \sim \frac{\hbar a}{\mu G m_{C}^2 \Delta t}
\label{tubnum}
\eeq

\section{Large Extra Spatial Dimensions}
\label{ledsection}
It is a long-standing supposition in theoretical physics that our Universe
might contain more that the traditional three spatial dimensions. 
The initial proposal made by Kaluza \cite{kaluza} and later by 
Klein \cite{klein} posited that a (4+1)-dimensional metric could help 
unify the fundamental
fields of gravitation and electromagnetism.  This extra dimension has 
traditionally assumed to be compactified with a small radius $R \ll 1$, 
which has until recently been taken to be of the Planck length ($>10^{-35}~$m).

The emergence of theories of large extra dimensions (LEDs) in the late
1990s relaxed the constraint that the size of any additional spatial dimensions
must be Planck order (see \cite{add,rs} for their pioneering works).  Although
introduced initially as a solution to the hierarchy problem, LED theories
have found numerous applications in the fields of particle physics and
cosmology, including candidates for missing energy in GeV-scale accelerator 
collisions \cite{paccel}, cosmic ray flux spectra anomalies \cite{rays}, and 
dark energy phenomenology \cite{uaccel}.  This manuscript will deal only with 
compactified Kaluza-Klein-like ADD extra dimensions \cite{add}, and not the
Randall-Sundrum ``warped'' dimensions \cite{rs}.

The energy scale (or temperature) at which gravity is expected to unify with 
the other fundamental
forces is exceedingly large, $M_{Pl} \sim 16~$TeV.  Such energies would have
been present a brief fraction of a second after the Big Bang (as a 
comparison, the present background energy of the Universe is in the range of
$10^{-4}~$eV).  The large energy scale in 
turn explains the smallness of the gravitational constant, since 
$G_N = M_{Pl}^{-2}$.
The underlying assumption of LED theory is that the electroweak unification
scale $M_{EW} \approx 1~$TeV is {\it also} the fundamental scale of gravitation
$M_{4+n}$ in the full $4+n$ dimensional spacetime.  Just as
the electroweak coupling goes as $m^{-2}_{EW}$, the ``actual'' 
gravitational coupling ($G_{4+n}$) is set by this new energy scale.

In $3+n$ spatial dimensions, the Newtonian potential 
differs from the familiar $1/r$ form.  If there are $n$ extra
compactified dimensions of scale size $R_n$, then for distances $r < R_n$ 
it can be shown ({\it e.g.} via Gauss' Law) that 
\beq
|\phi_n(r)| \sim \frac{G_{4+n} m}{r^{n+1}} ~,
\label{lednewton}
\eeq
(up to some geometric constants)
since the fields can now propagate in any of the $3+n$ spatial dimensions.
Gravitational interactions in the full $4+n$ dimensional spacetime are mediated
by the new coupling constant $G_{4+n}$, whose value is set by the TeV
unification scale.  For small $r$, the strength of gravitation becomes
commensurate with the other fundamental forces, particularly electromagnetism.

Gravity's apparent weakness at macroscopic distances arises from the
fact that the range of influence of the extra dimensions is limited to some
distance $R_n$, which is just the dimensions' compactification scale.
As a result, the value of the traditional Newtonian constant $G_N$ is
fixed by the size and number of the extra dimensions.

Continuity of the gravitational field at $r=R_n$ requires
\beq
\frac{G_{4+n} m}{R_n^{n+1}}  \sim  \frac{G_N m}{R_n}~~.
\eeq
So a relationship between the two coupling constants may be derived as
\beq
G_{4+n} \sim R_n^n G_N~~,
\label{g4n}
\eeq
which indicates that the value of the compactification radius is
\beq
R_n \sim 10^{\frac{32}{n}-19}~{\rm meters}
\eeq

As counter-intuitive as the result may seem, the traditional laws of 
Newtonian gravity might break down below this length scale.
It can quickly be shown that if large extra compactified dimensions exist, 
there must be more
than one.  If $n=1$, then $R\sim 10^{11}~$m, which would imply that 
deviations from Newtonian gravity should be observed at scales of the order
of the solar system (which clearly they do not).  
Probing the cases $n\geq 2$ has subsequently become a hot
topic of research, ranging from the aforementioned 
astrophysical observation and accelerator searches to high-precision
table-top laboratory experiments.  Via Cavendish-like gravitation 
experiments, the 
E\"{o}twash group \cite{eotwash} has determined that gravity behaves in
the classical Newtonian manner at scales as small 
as 200~$\mu$m, but a recent result has set this limit further down to 
100~nm \cite{decca}.

\section{Penrose-Di\'osi OR with LEDs}
\label{ledcollapse}
Before addressing the effect of possible LEDs on something so contrived
as the Orch-OR mechanism, their influence on basic quantum mechanics is
first addressed.  
In a recent paper \cite{ledor}, it was demonstrated that the existence of LEDs
can have measurable impact on nucleon collapse times described by Penrose's
orchestrated reduction paradigm\footnote{The idea that LEDs could influence
gravitation collapse schemes was anecdotally mentioned in Reference~\cite{tamas}.}

Assuming a nucleon wavefunction evolves such that the physical conformations
of two eigenstates are separated by a distance on par with the radius of the 
nucleon itself ($10^{-15}~$m), collapse times in the presence of LEDs become
much shorter than the $10^{15}~$seconds predicted in Reference~\cite{penrose1}.
In fact, if there are between $n=2$ and $3$ dimensions of compactification
scale $R_2 \sim 10^{-3}~$m to $R_3 \sim 10^{-9}~$m, the nucleon wavefunction
will collapse in under $10^{-5}~$seconds.  This short superposition time
would destroy the quantum mechanical nature of the nucleons, and thus
would have observable consequences for neutron diffraction.
Thus, these cases are ruled out by experiment.

If there are 4 or 5 dimensions of scale $10^{-11}~$m and $10^{-13}~$m
respectively, the collapse time increases to between 0.01-10~seconds.
These case again could easily be verified experimentally, and the result
could serve to constrain the LED mechanism (if Penrose's initial collapse
scheme is correct).  The cases $n=6$ to $n=8$ are of interest, because the
nucleon collapse times increase from $10^7~$seconds to $10^{15}~$seconds.
It would be difficult to test whether or not a nucleon may be superposed
for more than a year without succumbing to collapse, and thus it leaves
the question open as to whether or not these LED parameters are physically
viable.

\section{Orch-OR with LEDs}
\label{orchorled}
Since particles separated by distances less than the compactification radius 
of any LEDs will experience ``stronger'' gravity, there could be significant
consequences for the Orch-OR mechanism.  Modifying the Penrose-Hameroff
derivation from Section~\ref{penham}, the reduction is now calculated
using a potential function of the form in Equation~\ref{lednewton}.  

Equation~\ref{tubnum} can be modified to include LEDs by the replacement 
$G \longrightarrow G_{4+n}$, and $a \longrightarrow a^{n+1}$, giving
\beq
N_T \sim \frac{\hbar a^{n+1}}{\mu G_{4+n} m_c^2}
\eeq
Table~\ref{tab1} shows calculated values of $N_T$ for $n=2$ through 
$n=7$ extra dimensions.  The three neural frequencies mentioned in
Section~\ref{penham} are considered, which correspond
to collapse times of 25~ms, 100~ms, and 500~ms.  Additionally, a fourth
frequency is also included which represents a 5~ms neural signal specific
to human beings (see \cite{orchor})
requiring $10^{11}$ tubulin with regular Newtonian gravity.
The case $n=1$ is excluded because of the aforementioned astrophysical 
constraints, and $n\geq 8$ are also excluded because the size $R_n$ drops
below the mass separation (and thus regular Newtonian gravity would resume
at about this point).  Experimental tests of Newton's
inverse square law \cite{eotwash} have effectively ruled out extra 
dimensions above a few hundred microns ($10^{-4}~$m) in size, so it is 
also unlikely that $n=2$ is valid.

For the cases where $3 \leq n \leq 5$, a very startling result is observed.  
Since gravity is so much stronger than normal at distances $r \ll R_n$,
the self-energy of a single nucleon superposition is {\it larger} than
the total required energy $\Delta E$.  For instance, if there are $n=3$
extra dimensions whose scale size is roughly 1~nm, the ratio of the 
superposition self-energy of a single nucleon to the total energy required
for a 500~ms collapse would be $E_3/\Delta E \sim 10^6$.  The corresponding
number of tubulin required for the 100~ms and 25~ms scenarios are also 
unphysical for these cases.  Thus, if there are indeed 5 or less extra 
dimensions of the variety described by the ADD theory, then it is virtually 
impossible for this mechanism to be responsible for conscious 
correlates (as described by Penrose and Hameroff).  
However, as discussed in the previous section, these cases would also be
ruled out by experiment.

If there are seven or more extra dimensions, then their length scale size 
drops below
the separation of the carbon nuclei in the protein qubits, and one would
expect ``regular'' gravity to take over (as suggested by the reported data).
However, it is just below this value of $n$ that the implications of Orch-OR 
become striking.  In a universe with $n=6$ extra 
dimensions of scale 
$R_6 \sim 10^{-14}~$meters, the number of tubulin qubits $N_T$ becomes 
physically-realizable.  This number, however, is exceedingly small, 
on the order of a few hundred for the 500~ms
case. This number increases by order of magnitude with decreasing
$\Delta t$, giving $N_T \sim 10^3$ tubulin for 100~ms and $N_T \sim 10^4$ 
for 25~ms.  In this case, it would imply that effectively {\it all} 
biological organisms containing tubulin cytoskeletons are conscious, since 
as previously mentioned typical neuron contains $10^7$ 
tubulin!  Thus, even microscopic organisms with a relatively small number
of neurons would be conscious.  While philosophers might be extremely 
intrigued by this conclusion which opens the doors for a re-evaluation
of a being's self-awareness, the likelihood of this being a reality is,
to say the least, suspect.

\section{Variation of TeV scale}
The values in Table~\ref{tab1} assume that the unification scale is
$M_{4+n} \sim 1~$TeV.  However, there is nothing to suggest that it
cannot be slightly larger than this.  Table~\ref{tab2} demonstrates
how the values might change if $M_{4+n}$ shifted by a few orders of
magnitude and instead is $M_{4+n} \sim 10^{\delta}~$TeV.  The main
effect of raising the unification scale is to make the compactification
scale of dimensions smaller. 

Table~\ref{tab2} shows how the number of tubulin required for the
constant instances of Table~\ref{tab1} might change if $\delta = 2$ 
({\it i.e.} a unification scale of $100~$TeV).  Note that now only
the Orch-OR mechanism will be affected for only up to $n=4$ extra dimensions
before the compactification scale drops below the carbon nuclei superposition
separation $a$.  In fact, if there are 2 or 3 extra dimensions (1
is still ruled out by macroscopic gravitational phenomena) the OR
framework again cannot be the driving mechanism of state collapse.

\section{Evolution of consciousness and time dependent LEDs}
In Reference~\cite{orchor} a discussion of evolution and the emergence of
consciousness was raised, based on the estimation of $10^9$~tubulin
required for pre-conscious processing.  There is increasing observational
evidence that the value of Newton's constant $G$ has changed over the 
evolution of the Universe to present day (see References~\cite{marciano,loren}
for a pre-LED and post-LED discussion of time-dependent compactification
radii).  From Equation~\ref{g4n}, it can be deduced that a time-dependent constant
$G(t)$ varies as $R(t)^{-n}$.  So, depending on whether $G(t)$ is getting
bigger or smaller with time can be related to the changing scale size
of the extra dimensions.  

In fact, it can easily be shown that if $G(t)$
is getting bigger, then $R(t)$ must be getting smaller.  If the Orch-OR
mechanism is correct, then the implications for conscious emergence are
striking.  As has been shown in this analysis, large values of $R$ imply
either unphysical interpretations of Orch-OR, or alternatively that 
conscious processes require only a few microtubulin strands.  This could
suggest that pre-evolutionary microbes possessed conscious though 
(depending on the initial size of $R(t)$, that is).
Conversely, a shrinking value of $G(t)$ implies that the scale $R(t)$ is 
getting larger over time, implying that over large time scales more
organic entities will eventually achieve consciousness.

Of course, the time scales required for a significant change in the
value of $G(t)$ are on the order of a fraction of the age of the Universe,
which most likely would surpass the ``biological'' time of species on
the Earth.  Indeed, it has been shown that the compactification radii have
grown by less than a factor of ten in size over the history of the solar 
system \cite{ledor}.  Also, the variation of $G(t)$ is also independent of the
possible existence of extra dimensions.  Hence, the associated impact on
such quantum mechanical brain processes would still be relevant, and
thus opens intriguing speculation on how intelligence might have evolved 
elsewhere in the early Universe.

\section{Conclusions}
This article has examined the compatibility of Penrose and Hameroff's
orchestrated objective reduction model for consciousness in light of the
possible existence of large extra compactified spatial dimensions of the 
ADD variety.  Since the basis of the objective reduction model is explicitly
gravitational and Newtonian, the presence of LEDs and TeV gravity will 
significantly alter the conclusions drawn in Reference~\cite{orchor}.  
In fact, for extra dimensions of scale size less that $\sim 10^{-14}~$m
the Orch-OR model becomes an incomplete theory.  The required number of
tubulin to maintain the observed conscious frequencies are either outrageously
small, or even unphysical ($N_T < 1$). 

The greatest test of TeV-gravity and LEDs will begin in 2007 when the Large
Hadron Collider is brought
on-line.  The accelerator will problem energy scales well above the TeV boundary,
and as such will provide an exciting glimpse at a range of possible new
physics which exists in and beyond this energy range.  If extra dimensions
exist and are large compared to the Planck scale, their existence is expected
to be confirmed.  If they do exist, then the Orch-OR model is incomplete
or incorrect.  If they do not, then the mystery of consciousness and its
connection to quantum gravity is possibly one step closer to being explained.

\pagebreak
\begin{table}[h]
\begin{center}
\begin{tabular}{cc|cccc}\hline
&  & \multicolumn{4}{|c}{$N_T$} \\ 
$n$ &$R_n~$(meters)  & $T = 5~$ms & 25~ms & 100~ms & 500~ms \\ \hline
2 & $10^{-3}$  &$10^{-13}$& $10^{-14}$ & $10^{-14}$ & $10^{-20}$\\
3 & $10^{-9}$  &$10^{-9}$& $10^{-9}$  & $10^{-10}$ & $10^{-14}$\\
4 & $10^{-11}$ &$10^{-4}$& $10^{-5}$ & $10^{-6}$ & $10^{-6}$\\
5 & $10^{-13}$ &$2$& $0.5$ & $10^{-1}$ & $10^{-2}$\\
6 & $10^{-14}$ &$5\times 10^4$& $10^4$ &  $10^3$ & $500$\\
7 & $10^{-15}$ &$10^9$& $10^8$& $5\times 10^7$& $10^{7}$ \\ \hline
\end{tabular}
\label{tab1}
\caption{The number of tubulin proteins required for an orchestrated reduction
of $\Delta t = 5, 25, 100,$ and $500~$ms duration in Universes with $n$
extra dimensions.  The case $n=1$ is ruled out by the observed behavior of
macroscopic gravity, while the cases $n > 7$ would reproduce the ``standard''
Orch-OR results.  }
\end{center}
\end{table}

\begin{table}[h]
\begin{center}
\begin{tabular}{cc|cccc}\hline
 &  & \multicolumn{4}{|c}{$N_T$} \\
$n$ & $R_n~$(meters)& $T = 5~$ms & 25~ms & 100~ms & 500~ms \\ \hline
2 & $10^{-7}$  &$10^{-5}$& $10^{-6}$ & $6\times 10^{-7}$ & $10^{-7}$\\
3 & $10^{-12}$ &$30$     & $6$       & $1$               & $0.3$\\
4 & $10^{-14}$ &$10^8$   & $10^{7}$  & $10^{6}$          & $7\times 10^{5}$\\ \hline
\end{tabular}
\label{tab2}
\caption{The number of tubulin proteins required for an orchestrated reduction
of $\Delta t = 5, 25, 100,$ and $500~$ms duration in Universes with $n$
extra dimensions in which the gravitational unification scale is 100~TeV.
The size of each dimension $R_n$ is smaller than those in Table~1, and
thus regular Newtonian gravity is recovered at the tubulin length scales
for smaller $n$.}
\end{center}
\end{table}

\end{document}